\documentclass[a4paper]{jpconf}
\usepackage{graphicx}

\begin{document}

\title{$Z_2$-vortex order of frustrated Heisenberg antiferromagnets in two dimensions}

\author{Hikaru Kawamura}
\address{Department of Earth and Space Science, Faculty of Science,
Osaka University, Toyonaka 560-0043,
Japan}
\ead{kawamura@ess.sci.osaka-u.ac.jp}
\date{\today}
\begin{abstract}
We discuss the recent experimental data on various frustrated quasi-two-dimensional Heisenberg antiferromagnets from the viewpoint of the $Z_2$-vortex order, which include $S$=3/2 triangular-lattice antiferromagnet NaCrO$_2$, $S$=1  triangular-lattice antiferromagnet NiGa$_2$S$_4$, $S$=1/2 organic triangular-lattice antiferromagnets $\kappa$-(BEDT-TTF)$_2$Cu$_2$(CN)$_3$ and EtMe$_3$Sb[Pd(dmit)$_2$]$_2$, and $S$=1/2 kagome-lattice antiferromagnet volborthite Cu$_3$V$_2$O$_7$(OH)$_2\cdot$2H$_2$O, {\it etc\/}.
\end{abstract}

\section{Introduction}

 Recently, geometrically frustrated magnets have attracted much  interest because of their unconventional ordering behaviors, including  the possible quantum spin-liquid state \cite{Anderson,PLee,Balents} and the novel  ordered states like chiral ordered state \cite{MiyashitaShiba,Onoda,OkuboKawamura-chiral,Kawamura-SG}. Such novel ordered  states as well as the associated phase transitions are often borne by  novel excitations inherent to frustrated systems. A $Z_2$-vortex,  stabilized in a class of two-dimensional (2D)  Heisenberg magnets with the locally noncollinear spin order  \cite{KawamuraMiyashita}, is a typical example of such novel excitations inherent to geometrically frustrated  Heisenberg magnets. It was first identified in 1984 in the antiferromagnetic (AF) classical  Heisenberg model on the 2D triangular lattice as a topologically stable point  defect characterized by a two-valued topological quantum number \cite{KawamuraMiyashita}.  In this $Z_2$-vortex,  Heisenberg spins, or more precisely chirality vectors, circulate around a vortex core making a topologically stable vortex (Fig.1), whereas whether they circulate in clockwise or counter-clockwise fashion makes no distinction topologically. In other words, the $Z_2$-vortex has no characteristic winding number in sharp contrast to the standard vortex.

\begin{figure}[ht]
\begin{center}
\includegraphics[scale=0.4]{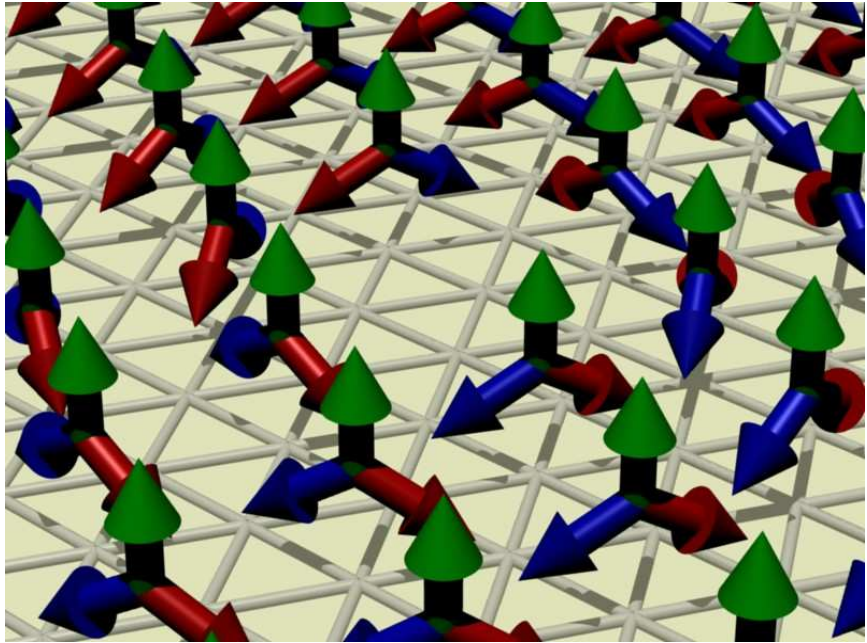}
\end{center}
\caption{
'`'Ž illustration of a $Z_2$-vortex.  Taken from [T. Okubo and H. Kawamura, J. Phys. Soc. Jpn. {\bf 79} (2010) 084706].
}
\end{figure}

 More recently, it was argued \cite{KawamuraYamamoto2,KawamuraYamamoto1,OkuboKawamura-vortex} that such a $Z_2$-vortex excitation might be responsible for certain anomalous behaviors observed experimentally in several triangular-lattice  Heisenberg AFs such as $S$=3/2 NaCrO$_2$  \cite{Olariu,Hsieh,Hsieh2,Hemmida}, $S$=1 NiGa$_2$S$_4$  \cite{Nakatsuji,Nambu,Takeya,Yaouanc,MacLaughlin,Yamaguchi,Yamaguchi2,Nakatsuji-review,Nambu2,Stock,Takubo} and $S$=1/2  organic compounds, $\kappa$-(BEDT-TTF)$_2$Cu$_2$(CN)$_3$ \cite{Shimizu,Kawamoto,Kurosaki,Shimizu2,Ohira,Yamashita,Yamashita2,Manna,Jawad} or EtMe$_3$Sb[Pd(dmit)$_2$]$_2$ \cite{Tamura,Itou,Itou2,Yamashita2-dmit}. In these materials, a spin-liquid-like  behavior without the standard magnetic long-range order (LRO)  was observed, while all of these compounds  exhibit a weak but clear transition-like anomaly at a  finite temperature.  As a possible explanation of such an experimental anomaly, a $Z_2$-vortex driven topological transition was invoked \cite{KawamuraYamamoto2,KawamuraYamamoto1,OkuboKawamura-vortex}. Although relatively little attention has been paid in the literature to $Z_2$-vortices  compared with, {\it e.g.\/}, the standard $Z$-vortex realized in  the two-dimensional {\it XY\/} model \cite{KT1,KT2}, it might have important  relevance to the ordering of a variety of frustrated quasi-2D magnets than hitherto thought generally. The aim of the present article is to point out such possible relevance of  $Z_2$-vortex excitations to the ordering process of various geometrically frustrated magnets under recent extensive study.

It was suggested in Ref.\cite{KawamuraMiyashita} that the frustrated 2D Heisenberg AF sustaining a $Z_2$-vortex might exhibit a thermodynamic phase transition at a finite temperature driven by the  binding-unbinding of the $Z_2$-vortices. The transition is of topological character where the ergodicity is broken topologically in the sense that the phase space is narrowed from the doubly connected one to the singly connected one.  An interesting observation is that the standard AF spin correlation length $\xi_s$ is kept finite even at and below the $Z_2$-vortex transition point $T=T_v$. Nontopological excitations such as spinwaves which survive below $T_v$ are enough to destroy the spin correlation to be paramagnetic with finite $\xi_s$. The low-temperature state characterized by the topologically broken ergodicity and by the finite (but often long) spin correlation length and correlation time is called a `spin-gel' state \cite{KawamuraYamamoto2}: See Fig.2. 

 Remarkable feature of the proposed $Z_2$-vortex transition might be the  decoupling of the two length scales: At the $Z_2$-vortex transition temperature  $T=T_v$, the vortex correlation length $\xi_v$ corresponding to the mean separation of free $Z_2$-vortices diverges, while the  spinwave correlation length $\xi_{sw}$ is kept finite (Fig.2(left)) \cite{KawamuraYamamoto2}. Reflecting the  finiteness of $\xi_{sw}$,  the full spin correlation length $\xi_s$ is kept finite even at $T < T_v$.   Possible decoupling  behaviour discussed above might be in sharp contrast to the behavior of another vortex-driven  transition, {\it i.e.\/}, the standard Kosterlitz-Thouless (KT) transition of  the 2D {\it XY\/} model, where there exists only one diverging length scale and the spin correlation length stays infinite throughout the low-temperature phase \cite{KT1,KT2}. Finiteness of the spin correlation length even below $T_v$ is a pronounced feature of the spin-gel state to be contrasted to the standard KT ordered state. 

 Finiteness of the spin correlation length $\xi_s$ at $T=T_v$ has important consequences on the response of the system against weak perturbative interactions coupled to the spin, {\it e.g.\/}, magnetic field, magnetic anisotropy and interplane coupling \cite{KawamuraYamamoto2,KawamuraYamamoto1}. In the situation where $\xi_s$ stays finite at the topological transition $T=T_v$, if the perturbation $A$ is sufficiently small satisfying $A\xi_s(T=T_v)^2 < k_BT_v$, the $Z_2$-vortex transition and the low-temperature `spin gel' state remains essentially the same as in the unperturbed ones: See Fig.2 for comparison of the two types of vortex transitions. The situation might be contrasted to that in the KT transition of the 2D {\it XY\/} model, where, reflecting the divergence of the spin correlation length, even an infinitesimal perturbation leads to the 3D magnetic long-range order in real magnets.

\begin{figure}[ht]
\begin{center}
\includegraphics[scale=0.6]{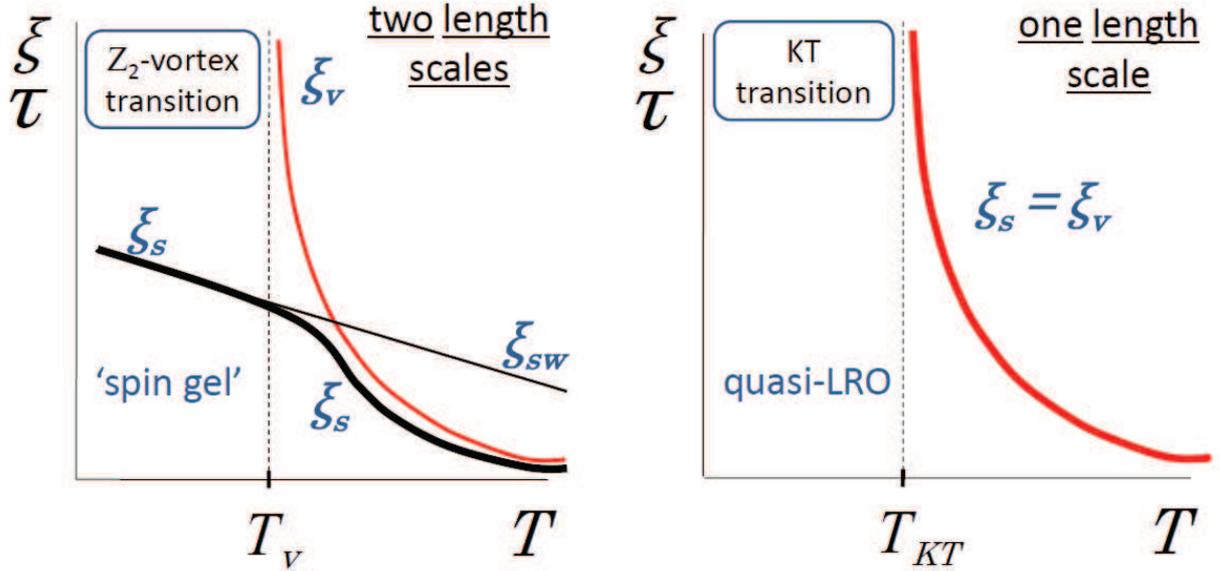}
\end{center}
\caption{
Schematic figure of the growth of the vortex correlation length $\xi_v$, the spinwave correlation length $\xi_{sw}$ and the full spin correlation length $\xi_s$ for each case of the $Z_2$-vortex transition of the frustrated Heisenberg model (left), and of the standard Kosterlitz-Thouless transition of the two-dimensional {\it XY\/} model (right).
}
\end{figure}

We note that the question of whether the decoupling of the two length scales, each associated with the topological and nontopological degrees of freedom, really occurs or not is still at issue. If there is no such decoupling, the vortex binding-unbinding phenomenon might eventually become a sharp crossover rather than a genuine thermodynamic transition \cite{APYoung}. Recent numerical simulations of the 2D triangular-lattice classical Heisenberg AF has indicated that the $Z_2$-vortex binding-unbinding phenomenon looks like a thermodynamic phase transition, at least at the length scale of $10^3$ lattice spacings \cite{KawamuraYamamoto2}. Hence, it could be a very sharp crossover with enough experimental relevance, if not to be a genuine phase transition. In the present paper, we wish to discuss experimental consequences of such a $Z_2$-vortex binding-unbinding phenomenon of the frustrated 2D Heisenberg magnets, not entering into the question of whether it is a genuine phase transition or a sharp crossover.

\section{The properties of the $Z_2$-vortex ordering}

In this section,  according to the recent theoretical analysis \cite{KawamuraYamamoto2}, we wish to summarize the expected properties of the $Z_2$-vortex and its binding-unbinding topological transition.

\medskip\noindent
1) When one approaches $T=T_v$ from above, the vortex correlation length $\xi_{v}$ diverges as
\begin{equation}
 \xi_v \sim \exp \left[\left(\frac{A}{T-T_v}\right)^\alpha\right],\ \ \ T >T_v,
\end{equation}
with $\alpha < 1$. The standard spin correlation length grows rapidly toward $T_v$, but remains finite even at and below $T=T_v$, exhibiting only a weak essential singularity at $T=T_v$: See Fig.2(left). 

\medskip\noindent
2) Singular part of the free energy around $T_v$ is given by $f_v\approx \xi_v^{-2}$. 

\medskip\noindent
3) The specific heat exhibits a weak essential singularity at $T=T_v$, while a non-singular peak or a hump appears slightly above $T_v$. The latter corresponds to the temperature where the maximum number of $Z_2$-vortex pairs dissociate giving rise to the maximum entropy change, while the transition temperature $T_v$ corresponds to the temperature where the first $Z_2$-vortex pair dissociates. 

\medskip\noindent
4) The magnetic susceptibility also shows a week essential singularity at $T_v$.

\medskip\noindent
5) The $Z_2$-vortex transition accompanies an anomaly in the spin dynamics. When ones approaches $T_v$ from above, the spin correlation time grows sharply toward $T_v$, but not truly diverges at $T=T_v$, remaining finite even at and below $T_v$.  Even below $T_v$,  large and slow fluctuations should remain (`spin-gel' state).

\medskip\noindent
6) The $Z_2$-vortex transition and the low-temperature spin-gel state are robust against sufficiently weak magnetic perturbations such as magnetic anisotropy, interplane coupling and applied magnetic fields. When these perturbations get larger beyond a critical value, which is determined by the spin correlation length at $T_v$, the system would change itself into the more standard ordered states such as the AF long-range ordered state or the frozen spin-glass state, instead of the spin-gel state.

\medskip\noindent
7) The dynamical spin structure $S(q,\omega)$ exhibits a characteristic central peak originated from free $Z_2$-vortices above $T=T_v$ \cite{OkuboKawamura-vortex}.   $S(q,\omega)$ also exhibits a characteristic intensity at finite $\omega$ along the zone boundary originated from $Z_2$-vortex pairs.

\section{Experiments on frustrated Heisenberg antiferromagnets in two dimensions}

 We now wish to discuss possible implications of the $Z_2$-vortex ordering scenario summarized in the previous section to recent experiments on several quasi-2D frustrated Heisenberg AFs.  Those include (a) $S$=3/2 triangular-lattice AF NaCrO$_2$, (b)  $S$=1 triangular-lattice AF NiGa$_2$S$_4$, (c)  $S$=1/2 organic triangular-lattice AFs $\kappa$-(BEDT-TTF)$_2$Cu$_2$(CN)$_3$ and EtMe$_3$Sb[Pd(dmit)$_2$]$_2$, (d)  $S$=1/2 kagome-lattice AF Cu$_3$V$_2$O$_7$(OH)$_2\cdot$2H$_2$O, and (e)  $S$=3/2 kagome-bilayer (or pyrochlore-slab) AF SrCr$_{8-x}$Ga$_{4+x}$O$_{19}$, {\it etc\/}.

\medskip\noindent
(a) NaCrO$_2$

 NaCrO$_2$ is a $S$=3/2 triangular-lattice Heisenberg AF with its Curie-Weiss temperature $T_{CW}\simeq$ 290K. This magnet does not exhibit the conventional AF LRO down to low temperature, while it exhibits a magnetic short-range order (SRO) characterized by an incommensurate wavevector close to the 120-degrees structure \cite{Olariu,Hsieh,Hsieh2,Hemmida,Soubeyroux}. The origin of the incommensurate spin structure is ascribed either to an easy-axis-type magnetic anisotropy or to a weak interplane coupling. A clear but rounded specific-heat peak is observed at $T_{peak}=41$K, whereas a transition-like dynamical anomaly is observed at $T_f\simeq 30$K, a temperature slightly below $T_{peak}$. At $T=T_f$, the spin dynamics is rapidly slowed down giving rise to a quasi-static internal field. The spin dynamics probed by  NMR, ESR and $\mu$SR, however, is not completely frozen below $T_f$. Rather, the spins remain slowly fluctuating even below $T_f$ unlike the conventional AF or the spin glass. Such a dynamically fluctuating ordered state extends over a wide temperature range down to 10K. The spin correlation length determined from neutron scattering is kept finite  $\xi\simeq 20$ lattice spacings  even below $T_f$, which is not resolution-limited.

 These experimental features are fully consistent, at least qualitatively, with the $Z_2$-vortex scenario discussed in the previous section, if the experimental freezing temperature $T_f$ is identified as $T_v$  and the low-temperature state as the `spin-gel' state. In particular, energetics seems appropriate. Recent Monte Carlo simulation on the triangular-lattice classical Heisenberg model with the nearest-neighbor AF coupling yielded $T_{peak}/T_{CW}\simeq 0.137$ and  $T_v/T_{CW}\simeq 0.123$ \cite{KawamuraYamamoto2}, which are close to the corresponding experimental values for NaCrO$_2$, {\it i.e.\/}, 0.14 and 0.10, respectively.  It remains to be seen whether the dynamical spin structure factor exhibits a characteristic behavior predicted as a fingerprint of the $Z_2$-vortex, the point 7 above \cite{OkuboKawamura-vortex}.  

 Magnetic ordering of similar chromium triangular-lattice AFs like HCrO$_2$ \cite{Hemmida,Ajiro}, LiCrO$_2$ \cite{Hemmida,Ajiro,Soubeyroux,Kadowaki-LiCrO2,Alexander,Tokura,Olariu-LiCrO2}, CuCrO$_2$ \cite{Tokura,Kadowaki,Poienar,Kimura},  AgCrO$_2$ \cite{Tokura,Oohara} and  PdCrO$_2$ \cite{Mekata-PdCrO2,Takatsu} were also investigated experimentally. In contrast to NaCrO$_2$, these materials exhibit a 3D AF LRO at a Neel temperature $T=T_N$ due to the weak interlayer coupling. Yet, since all these materials are magnetically quasi-2D, the $Z_2$-vortex might exist in the 2D regime realized in the temperature region slightly above $T_N$. A particularly interesting compound in this series might be KCrO$_2$, where no evidence of the 3D AF LRO was reported by neutron scattering as for NaCrO$_2$ \cite{Soubeyroux}. Further experimental study on this compound is encouraged.

 It was also reported that the two-dimensionality could strongly be enhanced in Cu$_{1-x}$Ag$_x$CrO$_2$, a compound obtained from  CuCrO$_2$ by chemical substitution \cite{Okuda,Kajimoto}. Then, one might expect that the $Z_2$-vortex has a higher chance to be realized in Cu$_{1-x}$Ag$_x$CrO$_2$. Of course, in this compound, care has to be taken on the possible effect of quenched randomness.

\medskip\noindent
(b) NiGa$_2$S$_4$ 

  NiGa$_2$S$_4$  is a $S$=1 triangular-lattice Heisenberg AF with its Curie-Weiss temperature $T_{CW}\simeq$ 80K \cite{Nakatsuji,Nambu,Takeya,Yaouanc,MacLaughlin,Yamaguchi,Yamaguchi2,Nakatsuji-review,Nambu2,Stock,Takubo}. This material attracted considerable attention of researchers, because the low-temperature specific heat exhibits a $T^2$ behavior suggesting the existence of Goldstone modes associated with a broken continuous symmetry, whereas no magnetic LRO was observed down to low temperature \cite{Nakatsuji}. The observed magnetic SRO is an incommensurate helical structure in the basal plane close to the 60-degrees structure. Such a spin structure is stabilized by the AF third-neighbor interaction $J_3$ which frustrates the ferromagnetic first-neighbor coupling $J_1$ with $J_1/J_3\simeq -0.2$ \cite{Nakatsuji-review}.

 Meanwhile, NiGa$_2$S$_4$, with an integer spin quantum number $S$=1 in contrast to a half-integer spin quantum number $S$=3/2 of NaCrO$_2$, exhibits a strikingly similar ordering behavior to that of NaCrO$_2$. Indeed, NiGa$_2$S$_4$ exhibits a clear but rounded specific-heat peak at $T_{peak}=12$K \cite{Nakatsuji-review} (the peak of $C/T$ is located at 10K), whereas a transition-like dynamical anomaly is observed at $T^*\simeq 8.5$K, a temperature slightly below $T_{peak}$, where the spin dynamics is rapidly slowed down. Note that quasi-static internal fields of {\it dipolar\/} nature is observed below $T^*$, at least at the time scale of NQR and muon measurements. The spin dynamics probed by  NMR, NQR, ESR and $\mu$SR, however, is not completely frozen even below $T^*$. Rather, the spins remain slowly fluctuating at the time scale of MHz, unlike the conventional AF or the spin glass. Such a dynamically fluctuating ordered state extends over a wide temperature range down to 2K. The spin correlation length determined from neutron scattering is kept finite even at and below $T^*$, {\it i.e.\/}, $\xi\simeq 7$ lattice spacings \cite{Nakatsuji}. Dynamical freezing at $T=T^*$ also accompanies a weak anomaly in the susceptibility. Anomaly in the susceptibility looks clearer than the one observed in NaCrO$_2$, presumably due to the fact that NiGa$_2$S$_4$ possesses a ferromagnetic nearest-neighbor coupling. Application of magnetic fields greater than $100\sim 1000$ G gradually changes the slowly fluctuating state into the more conventional frozen state \cite{MacLaughlin}.

 As for NaCrO$_2$, those experimental features of  NiGa$_2$S$_4$ are fully consistent, at least qualitatively, with the $Z_2$-vortex ordering scenario, if the experimental freezing temperature $T^*$ is identified  as $T_v$ and the low-temperature state as the spin-gel state. Concerning its energetics, NiGa$_2$S$_4$  is characterized by $T_{peak}/T_{CW}\simeq 0.15$ and  $T_v/T_{CW}\simeq 0.11$, which are close to the corresponding theoretical values 0.133 and 0.123 \cite{KawamuraYamamoto2}. Furthermore, a theoretical estimate of the crossover-field intensity $\sim 1000$G obtained with the information of the spin correlation length $\xi\simeq 6$ lattice spacings at $T=T^*$, seems consistent with the $\mu$SR measurement \cite{MacLaughlin}.

  In this $Z_2$-vortex picture, slow dynamics below $T^*$ is primarily borne by spinwaves. Then, spinwaves would be responsible for the $T^2$ specific heat. Indeed, recent inelastic neutron-scattering measurements identified a damped spinwave excitation at a low temperature $T=1.5$K \cite{Stock}. The $T^2$ behavior of the specific heat sets in around $T^*$, which is consistent with such a picture. Ref.\cite{Fujimoto} accounted for the $T^2$ specific heat based on spinwave excitations of the noncollinear AF order of $S$=1 quantum magnets, neglecting the vortex degrees of freedom \cite{Fujimoto}. Vortex-free assumption of Ref.\cite{Fujimoto} is well justified below the $Z_2$-vortex transition. 

 Impurity effects on the low-temperature specific heat and on the transition-like anomaly $T^*$ was studied in Ref.\cite{Nambu2}. Upon substitution of various kinds of impurities, {\it i.e.\/}, $S=0$ Zn$^{2+}$, $S=2$ Fe$^{2+}$,  $S=3/2$ Co$^{2+}$ and  $S=5/2$ Mn$^{2+}$, the transition-like anomaly of the pure compound at $T^*$  gradually changed its character to that of the standard spin-glass transition in all cases studied, accompanied by a pronounced cusp and the deviation between the FC and ZFC data. By contrast, the low-temperature specific heat persistently exhibited a $T^2$ behavior scaled by the Curie-Weiss temperature for Zn$^{2+}$ and Fe$^{2+}$ with integer spins, but exhibited a different $T$-linear behavior characteristic of the standard spin glass for Co$^{2+}$ and Mn$^{2+}$ with half-integer spins. This observation seems to suggest that the nature of low-energy excitations depends significantly on the size of spins, {\it i.e.\/}, whether $S$ being an integer or a half-integer. It remains to be seen whether the standard spinwave picture could account for such exotic properties or not. In this context, one may also notice that, in case of kagome-bilayer AF SrCr$_{8-x}$Ga$_{4+x}$O$_{19}$ (SCGO) to be discussed below, the low-temperature specific heat is observed to persistently exhibit a $T^2$ behavior below the spin-glass transition temperature, although the spin quantum number there is a half-integer ($S=3/2$ for Cr$^{3+}$): See section 3(e) below.

 Concerning the magnitude of the spin correlation length,  quantitative discrepancy exists between the experimental value and the numerical value obtained for the simplest classical Heisenberg model with the nearest-neighbor AF coupling. Monte Carlo simulation yielded $\xi$ of order $10^3$ near $T=T_v$ \cite{KawamuraYamamoto2}, while it is only 8 or 20 experimentally. So, some mechanism, not taken into account in the simplest classical Heisenberg model, is required to explain the shortness of experimental $\xi$. It should be emphasized here that the magnitude of $\xi$ is a nonuniversal property governed by the non-vortex physics, which could significantly be reduced, say, by quantum fluctuations, charge fluctuations, or further frustration effects associated with the interaction other than the main exchange couplings. Interestingly, recent ARPES study performed at higher temperatures (100$\sim$200K) has revealed that the low-energy charge fluctuation (hole dynamics) across the Mott gap is characterized by the wavevectors which are completely different from the wavevectors characterizing the mangnetic SRO at lower temperatures, the occurrence of spin-charge separation \cite{Takubo}.

 In fact, the specific heat of NiGa$_2$S$_4$ exhibits another pronounced peak at a temperature around $T\sim 100$K, an order of magnitude higher temperature than that at $T_{peak}=10$K. This higher specific-heat peak might be related to charge fluctuations as investigated in Ref.\cite{Takubo}. Ref.\cite{Balents2} (see also Ref.\cite{Takubo}) suggested that the strong nematic correlations might set in around this higher specific-heat-peak temperature. However, the $T$-linear behavior in the specific heat mentioned above sets in at the lower specific-heat-peak temperature of 10K, not at the higher specific-heat-peak temperatureof of 100K, becoming vanishingly small (after the lattice contribution subtracted) in the temperature range between the two specific-heat-peak temperatures. Hence, strong nematic correlations, as are expected to accompany nematic-wave excitations giving rise to the $T^2$ specific heat, are unlikely to occur in the high-temperature regime of $\sim$100K.

 Finally, the noncollinear AF order might explain another noticeable feature of experiments that the $T^2$ specific heat is quite robust against applied magnetic fields \cite{Nakatsuji}. This experimental observation is rather surprising since applied fields reduce the Hamiltonian symmetry from O(3) to O(2), leading to smaller number of Goldstone modes, {\it i.e.\/}, from three to one. As discussed in Ref.\cite{KawaMiya2},  an interesting observation here is that the noncollinear AF ground state in magnetic fields is capable of keeping an ``accidental'' degeneracy not related to the Hamiltonian symmetry O(2), essentially of  the same amount as in the zero-field case, when the ordered state is a commensurate one, {\it e.g.\/}, the 120-degrees or the 60-degrees structure. In fact, even in applied fields, the ground-state manifold still retains {\it three\/} continuous parameters which can be set freely, just as in the zero-field case. One of these three is of symmetry origin, {\it i.e.\/} a true Goldstone mode, while other two are not of symmetry origin (accidental), {\it i.e.\/ pseudo}-Goldstone modes. Thus, at the classical level, such pseudo-Goldstone modes may account for the robustness of the low-temperature specific heat in applied fields, while this degeneracy would become approximate in quantum systems. The degeneracy would also become approximate when the ordering wavevector deviates from the commensurate position.

\medskip\noindent
(c) $\kappa$-(BEDT-TTF)$_2$Cu$_2$(CN)$_3$ and EtMe$_3$Sb[Pd(dmit)$_2$]$_2$ 

 $S$=1/2 organic triangular-lattice AFs $\kappa$-(BEDT-TTF)$_2$Cu$_2$(CN)$_3$ \cite{Shimizu,Kawamoto,Kurosaki,Shimizu2,Ohira,Yamashita,Yamashita2,Manna,Jawad} and EtMe$_3$Sb[Pd(dmit)$_2$]$_2$ \cite{Tamura,Itou,Itou2,Yamashita2-dmit} exhibit a spin-liquid behavior without the conventional magnetic LRO down to low temperature. The lattice here is not a regular triangular lattice, but is a slightly distorted one. The charge and the spin densities are spread over a dimer molecule so that the system intrinsically possesses the polarization degree of freedom within a molecule. Furthermore,  these materials are Mott insulators lying close to the metal-insulator boundary so that intermolecular charge fluctuations might play an important role. 

 Recent measurements have revealed that the specific heat \cite{Yamashita} and the thermal conductivity \cite{Yamashita2-dmit} of these organic compounds exhibit a $T$-linear behavior at low temperatures, often ascribed to the fermionic `spinon'-like excitations \cite{PLee-spinon1,PLee-spinon2}. More precisely, for the case of $\kappa$-(BEDT-TTF)$_2$Cu$_2$(CN)$_3$, such a $T$-linear behavior was reported for the specific heat \cite{Yamashita}, while a small nonzero energy gap was reported for the thermal conductivity \cite{Yamashita2}. For EtMe$_3$Sb[Pd(dmit)$_2$]$_2$, even the thermal conductivity exhibits a $T$-linear behavior \cite{Yamashita2-dmit}.  

 Recent measurements have also revealed a weak transition-like anomaly occurring at a finite temperature just above the temperature range where the $T$-linear behavior is observed in the specific heat or in the thermal conductivity. In case of $\kappa$-(BEDT-TTF)$_2$Cu$_2$(CN)$_3$, two anomalies (or crossovers) were observed, one at $\sim$6K and the other at $\sim$3K. The higher one at 6K is characterized by the hump (or the peak) of the specific heat \cite{Yamashita} and of the thermal expansivity coefficient \cite{Manna}, with a minimum of the NMR $T_1^{-1}$ \cite{Shimizu} and of the in-plane thermal conductivity \cite{Yamashita2}. The lower one at 3K is characterized by a weak additional structure in the specific heat \cite{Yamashita} and in the thermal expansivity coefficient \cite{Manna}, with a maximum of the in-plane thermal conductivity \cite{Yamashita2}. Similar anomaly or crossover is also observed for EtMe$_3$Sb[Pd(dmit)$_2$]$_2$ \cite{Itou,Itou2,Yamashita2-dmit}.

 We note that the $Z_2$-vortex order is one possible candidate of the experimentally observed anomaly. In this correspondence, one identifies the lower anomaly as the $Z_2$-vortex transition temperature $T_v$ and the higher anomaly as the specific-heat-peak temperature $T_{peak}$. Experimentally, the ratio $T_{peak}/T_v$ comes around two in these organic compounds, which deviates somewhat from the values for NaCrO$_2$ and NiGa$_2$S$_4$ discussed above.

 Of course, for the vortex to be a meaningful excitation, minimal amount of noncollinear spin SRO, at least of a few lattice spacings, is required, which, however, is a plausible situation in these compounds in view of the observed spectral broadening of NMR signal and the gapless behavior of the susceptibility. 

 $\kappa$-(BEDT-TTF)$_2$Cu$_2$(CN)$_3$ exhibits a dielectric anomaly  in the temperature range 20K$\sim $40K \cite{Jawad}, somewhat above the temperature where the transition-like anomaly is observed  in the specific heat or in NMR $T_1^{-1}$. The AC dielectric constant exhibits a strongly frequency-dependent peak  at the time scale of kHz in this temperature range. The observed eminent frequency dependence then suggests that the polarization degree of freedom becomes slowed down significantly in this temperature regime and tends to be frozen. It remains to be seen how this slowing down or freezing of the polarization degree of freedom is related or unrelated to the spin-liquid and the transition-like behaviors observed at lower temperatures. An extrapolation of the $\omega$-dependent peak temperature to $\omega \rightarrow 0$ yields an estimate of the static dielectric transition temperature $T_c\sim 6$K, which happens to come close to the specific-heat hump temperature. On the basis of such an observation, Ref.\cite{Jawad} argued that the anomaly of the dielectric constant might be linked to the anomaly of the specific heat \cite{Jawad}. However, such coincidence should be taken with care, since the dielectric measurements are performed already at the macroscopic time scale ($\sim $kHz order) not much different from the time scale of thermal measurements and the observed peak temperature of the AC dielectric constant (20K$\sim $40K) is still considerably higher than the specific-heat hump temperature (6K). Thus, at the time scale of real measurements, the specific-heat hump is likely to be a phenomenon not directly related to the dielectric anomaly.

 The $T$-linear specific heat and of the thermal conductivity observed at  low temperatures cannot simply be explained by spinwaves. Some other excitations different from spinwaves or any type of Goldstone modes are definitely required here. Two possibilities may be given here: One is a fermionic `spinon' excitation possibly realized in the RVB-like quantum spin-liquid state, as extensively discussed in the literature \cite{PLee-spinon1,PLee-spinon2}. The other possibility I wish to propose here is a more conservative one.  As discussed above,
 since the polarization within a molecule has been dynamically frozen at temperatures higher than the transition temperature, and since the spin is likely to be coupled to the polarization in some way or other, randomly frozen polarization might act as a quenched randomness to the spin dynamics, which might give rise to low-energy spin excitations with a constant density of states at low energies, as often encountered in random spin systems.  In this context, it might be intereseting to point out that Ref.\cite{Shimizu2} observed that applied magnetic fields induced inhomogenous magnetization in $\kappa$-(BEDT-TTF)$_2$Cu$_2$(CN)$_3$, which was ascribed to impurities or defects. Of course, one needs to explain the reason why the spin itself escapes the freezing in such a situation, in contrast to the standard spin-glass case. Further study is required to clarify the nature of low-energy excitations relevant to the low-temperature properties of these compounds. 

 Anyway, one should recall that the issue of the low-energy excitations responsible for the low-temperature behavior could more or less be independent of the issue of a finite-temperature transition. Within the $Z_2$-vortex ordering picture, in particular, any type of excitation possibly realized in the vortex-free sector in the phase space, can be a candidate of relevant low-energy excitation.

\medskip\noindent
(d) Cu$_3$V$_2$O$_7$(OH)$_2\cdot$2H$_2$O

Volbothite Cu$_3$V$_2$O$_7$(OH)$_2\cdot$ 2H$_2$O is a $S$=1/2 Heisenberg AF on the kagome lattice \cite{H-Yoshida,M-Yoshida,Yamashita-volbo,Nilsen,Hiroi,Fukaya,BertBono,BertBono2}. In this compound, the lattice is slightly distorted in a uniaxial manner, {\it i.e.\/}, a regular triangle being distorted to an isosceles triangle. The extent of the distortion, however, seems not so large. Vesignieite BaCu$_3$V$_2$O$_8$(OH)$_2$ is also a  $S$=1/2 kagome-lattice Heisenberg AF with a small lattice distortion \cite{Okamoto,Zhang}. In vesignieite, the lattice distortion is even less than in volborthite.

 Volbothite also exhibits a spin-liquid-like behavior in the sense that no AF LRO is observed down to very low temperature \cite{H-Yoshida,M-Yoshida,Yamashita-volbo}. Recent neutron scattering experiment has revealed the onset of the magnetic short-range order at low temperatures \cite{Nilsen}.
Interestingly,  NMR and specific heat measurements have revealed that volborthite exhibits a transition-like anomaly at a finite temperature \cite{BertBono,H-Yoshida,M-Yoshida,Yamashita-volbo,Hiroi-pc}. Indeed, the NMR relaxation rate $T_1^{-1}$ exhibits a sharp peak suggestive of a thermodynamic  phase transition at around $T=T^*=0.9$K \cite{H-Yoshida,M-Yoshida}, while the specific heat shows a hump or a kink at $T_K=1.05$K \cite{Yamashita-volbo,Hiroi-pc}. Interestingly, a $T$-linear behavior of the specific heat is observed below $T^*$ together with the $T^2$-term, in spite of the fact that volborthite is a very good insulator. Similar behavior, {\it i.e.\/}, a combination of the $T$-linear term and the $T^2$ term, is observed also above $T^*$, but with different coefficients from those at $T<T^*$. A closer look at the data reveals that the peak temperature of NMR $T_1^{-1}$ and the kink (hump) temperature of the specific heat are not the same, the former $T=0.90$K being about 15\% lower than the latter $T=1.05$K. Such $\sim 15$\% difference reminds us of the $T_v$ vs. $T_{peak}$ relation associated with the $Z_2$-vortex ordering.

 Below $T^*$, spins remain fluctuating but the dynamics becomes very slow, as evidenced by NMR $T_2^{-1}$ measurements \cite{M-Yoshida}. (Volbothite also exhibits a weak spin-glass-like freezing phenomenon at a temperature different from $T^*$, which, however, is of extrinsic origin \cite{H-Yoshida,M-Yoshida}.) This slowly fluctuating low-temperature phase remains stable against weak magnetic fields. When an applied field intensity exceeds a critical value of $\sim 4.3$T, the system exhibits a phase transition into the more conventional AF long-range ordered state \cite{H-Yoshida,M-Yoshida}.

 Vesignieite also exhibits a spin-liquid-like behavior at low temperatures \cite{Okamoto,Zhang}. Although fuller experimental analysis is yet to be done, smaller lattice distortion of vesignieite makes this material a quite attractive model compound.

 Many of the above features of volborthite, {\it i.e.\/}, [A] the onset of a sharp dynamical anomaly at a finite temperature $T^*$, [B] the specific-heat hump (kink) temperature located slightly above ($\sim 15$\%) the dynamical transition temperature $T^*$, [C] appearance of a slowly fluctuating low-temperature phase without the conventional AF nor spin-glass LRO at $T<T^*$, and [D] the fluctuating low-temperature state stabilized against weak applied fields, which is transformed into the conventional LR ordered state upon application of stronger fields, are all quite similar to the features observed in triangular-lattice Heisenberg AFs NaCrO$_2$ and NiGa$_2$S$_4$ discussed above, and are also fully consistent with the $Z_2$-vortex ordering picture. 

 The existence of the $T$-linear term in the low-temperature specific heat is a feature  absent in  NaCrO$_2$ and NiGa$_2$S$_4$ (though present in organic triangular AFs).  The behavior of the low-temperature specific heat reflects the properties of the low-energy excitation of the system. If the relevant low-energy exciations are Goldstone modes like spinwaves, a $T^2$ term should arise as is observed in NaCrO$_2$ and NiGa$_2$S$_4$. Hence, the existence of the $T$-linear term indicates that the relevant low-energy excitation  in volborthite is exotic. Indeed, Kagome AF has been known to sustain exotic localized low-energy excitations like the weathervane excitation, which might lead to the observed $T$-linear specific heat. 

 It should be noticed again, however, that the properties of the low-temperature phase, {\it e.g.\/}, the existence or nonexistence of the $T$-linear term in the low-temperate specific heat, is a property governed by the non-vortex physics, an issue independent of the $Z_2$-vortex transition itself. In other words, the $Z_2$-vortex ordering scenario is well compatible with various different scenarios of the low-temperature phase, as long as the relevant low-energy excitations are realizable in a restricted vortex-free (topologically simply-connected) phase space.

 The other aspect to be considered is the role played by the Dzaloshinskii-Moriya (DM) interaction, which inevitably exists to certain extent in any kagome-lattice magnet. The DM interaction tends to lower the symmetry, lift the degeneracy of states, and thereby favor the ordering (usually favors the $q=0$ state). Thus, its effect on the nature of low-energy excitations needs to be further examined.

\medskip\noindent
(e) SrCr$_{8-x}$Ga$_{4+x}$O$_{19}$

SrCr$_{8-x}$Ga$_{4+x}$O$_{19}$ (SCGO) is a $S$=3/2 kagome-bilayer (or pyrochlore slab) system, extensively studied in 1990's as a typical model system of kagome-lattice Heisenberg AF \cite{Ramirez,Broholm,Martinez,Martinez2,Uemura,Lee,Keren,Ramirez2,Mendels,Mekata,Limot}. Crystal structure of magnetic ions (Cr$^{3+}$) in this material, however, is not really a planar kagome, but rather a kagome-bilayer where two kagome layers are connected via a sparse triangular layer in between, {\it i.e.\/}, a pyrochlore slab along (111) \cite{KawamuraArimori,ArimoriKawamura}. Thus, a constituent unit of the lattice is a tetrahedron rather than a triangle. Another point to be noticed is that a certain fraction of Cr ions are randomly diluted by nonmagnetic Ga so that the lattice contains sizable amount of disorder, which might affect the magnetic ordering of this material to some extent.

 SCGO exhibits a rather sharp spin-glass-like transition at a finite temperature $T_f$ characterized by a cusp in the susceptibility accompanied by the deviation between the FC and ZFC data. Even the negative divergence of the nonlinear susceptibility, which is a characteristic of the standard spin-glass order, was observed \cite{Ramirez}. This spin-glass-like transition, however, is unusual in the sense that (i) only a fraction of spin is frozen in the low-temperature phase and large spin fluctuations remain even below  $T_f$, (ii) the low-temperature specific heat is proportional to $T^2$ suggestive of 2D spinwaves (or certain Goldstone modes) in sharp contrast to the standard spin-glass case where the low-temperature specific heat exhibits a $T$-linear behavior, and (iii) the $T_f$-value of SCGO tends to increase with decreasing the defect concentration, apparently suggesting that the defect might not be essential in the spin-glass-like transition of SCGO \cite{Mekata}. The specific heat exhibits a rounded peak slightly above $T_f$ without an appreciable anomaly at $T=T_f$, as commonly observed in standard spin glasses.  Very much similar behavior was also reported for other $S$=3/2 kagome-bilayer compound Ba$_2$ Sn$_2$ZnCr$_{7p}$Ga$_{10-7p}$O$_{22}$ (BSZCGO) \cite{Hagemann,Bono,Bono2,Bonnet}.

 Some of the above features of SCGO, {\it e.g.\/}, considerable fluctuations remaining in the ordered phase and the specific-heat peak slightly above $T_f$, seem common with the features observed in other frustrated magnets discussed above, NaCrO$_2$, NiGa$_2$S$_4$ and volborthite. The occurrence of a clear spin-glass transition in SCGO, accompanied by the divergence of the nonlinear susceptibility, is at odds with other magnets, however. The implication from the observation (iii) above might be the spin-glass transition of SCGO is not intrinsically defect-mediated.

 One possibility in the context of the $Z_2$-vortex ordering scenario might be that the spin-glass transition of SCGO is essentially a $Z_2$-vortex transition, which takes a character of the spin-glass transition due to the presence of significant amount of defects (randomness). In other words, the $Z_2$-vortex transition serves as a ``seed'' of the experimentally observed spin-glass transition. In view of the fact that the observed spin-glass transition is quite pronounced accompanying the associated critical phenomena, it remains to be seen what role the $Z_2$-vortex is playing here. The true nature of the spin-glass transition of SCGO has remained elusive yet.

\section{Summary and discussion}

We discussed the recent experimental data on various frustrated quasi-2D Heisenberg AFs from the viewpoint of the $Z_2$-vortex order, which include (a) $S$=3/2 triangular-lattice AF NaCrO$_2$, (b) $S$=1  triangular-lattice AF NaGa$_2$S$_4$, (c) $S$=1/2 organic triangular-lattice AFs $\kappa$-(BEDT-TTF)$_2$Cu$_2$(CN)$_3$ and EtMe$_3$Sb[Pd(dmit)$_2$]$_2$, (d) $S$=1/2 kagome-lattice AF volborthite Cu$_3$V$_2$O$_7$(OH)$_2\cdot$2H$_2$O, and (e) $S$=3/2 kagome-bilayer AF  SrCr$_{8-x}$Ga$_{4+x}$O$_{19}$.

 These magnets exhibit the spin-liquid-like behavior without the conventional AF LRO at low temperatures. Furthermore, most of them exhibit the following features in common: [A] onset of a sharp dynamical anomaly at a finite temperature $T^*$ without a strong anomaly in the specific heat, [B] the specific-heat exhibiting a hump (or a kink) slightly above the dynamical transition temperature $T^*$, [C] appearance of a slowly fluctuating low-temperature state without the conventional AF nor spin-glass order at $T<T^*$, and [D] the fluctuating low-temperature state stabilized against weak applied fields, which is transformed into the conventional LR ordered state upon application of stronger fields. These features are observed in common at least for [a] NaCrO$_2$, [b] NiGa$_2$S$_4$ and [d] volborthite. Note that these materials are already of quite a variety, either $S$=3/2, 1 or 1/2, and either triangular or (distorted) kagome. Hence, the above features [A]-[D] should possess universal origin independent of the spin quantum number and the detailed lattice structure. The $Z_2$-vortex ordering scenario meets such a universal criterion since the conditions required there are, (i) the system should be 2D, {\it i.e.\/}, weak enough interplane coupling, (ii) the system should be Heisenberg-like, {\it i.e.\/}, weak enough magnetic anisotropy, (iii) the existence of the noncollinear spin SRO arising from frustration. All compounds discussed above satisfy these criteria to certain extent, and could be candidates of the $Z_2$-vortex bearing system. Such universal features of the $Z_2$-vortex ordering scenario is a favorable factor in explaining why the exotic properties [A]-[D] are observed in common in a variety of 2D frustrated Heisenberg AFs with different lattice geometries and spin quantum numbers. 

 $S$=1/2 organic compounds $\kappa$-(BEDT-TTF)$_2$Cu$_2$(CN)$_3$ and EtMe$_3$Sb[Pd(dmit)$_2$]$_2$ exhibit somewhat different properties from NaCrO$_2$, NiGa$_2$S$_4$ and volborthite discussed above. For example, $T_{peak}/T^*$-value takes a larger value of $\sim 2$. Dynamical slowing down in the low-temperature phase below $T^*$ seems less pronounced as compared with the one in the other compounds. As discussed in \S 3(c), the latter property as well as the $T$-linear low-temperature specific heat might reflect a novel type of low-energy excitations inherent to these compounds, but may well be compatible with the $Z_2$-vortex order.  Even for $S$=3/2 bilayer-kagome magnet SCGO, $Z_2$-vortex might be playing some role. The open issue here is to clarify the nature of the observed spin-glass-like  transition.

 Overall, the $Z_2$-vortex ordering scenario seems to give consistent account of a number  of  exotic ordering properties observed in a variety of frustrated quasi-2D Heisenberg magnets in common. Hopefully, further experimental and theoretical studies such as  inelastic neutron-scattering will reveal further novel features of the $Z_2$-vortex and other exotic excitations inherent to frustrated magnets.

\medskip

Special thanks are due to  T. Okubo for useful discussion and collaboration. Enlightening discussion with Z. Hiroi, S. Nakatsuji, Y. Nambu, N. Hagiwara, H. Yamaguchi,  H. Ohta, S. Maegawa, T. Itou, H. Kikuchi, Y. Okamoto, C. Broholm, D. MacLaughlin, P. Mendels, R. Kajimoto, K. Tomiyasu, M. Matsuura, Y. Motome, R. Kato, K. Kanoda, Y. Nakazawa, M. Yamashita, C. Hotta and T. Arima is gratefully acknowledeged. This work is supported by Grand-in-Aid for scientific Research on Priority Areas ``Novel States of Matter Induced by Frustration'' (19052006).


\bigskip
\bigskip


\begin{thebibliography}{00}

\bibitem{Anderson} Anderson P W 1973 Mater. Res. Bull. {\bf 8} 153
\bibitem{PLee}  Lee P A 2008 Science {\bf 321} 1306
\bibitem{Balents}  Balents L 2010 Nature {\bf 464} 199
\bibitem{MiyashitaShiba} Miyashita S and Shiba H 1984 J. Phys. Soc. Jpn. {\bf 53} 1145
\bibitem{Onoda} Onoda S and Nagaosa N  2007 Phys. Rev. Lett. {\bf 99} 027206.
\bibitem{OkuboKawamura-chiral} Okubo T and Kawamura H  2010 Phys. Rev. B {\bf 99} 014404
\bibitem{Kawamura-SG} Kawamura H 2010 J. Phys. Soc. Jpn. {\bf 79} 011007
 \bibitem{KawamuraMiyashita} Kawamura H and Miyashita S 1984  J. Phys. Soc. Jpn. {\bf 53}  4138.
\bibitem{KawamuraYamamoto2} Kawamura H, Yamamoto A and Okubo T, 2010 J. Phys. Soc. Jpn. {\bf 79} 023701
\bibitem{KawamuraYamamoto1} Kawamura H and  Yamamoto A 2007  J. Phys. Soc. Jpn. {\bf 76} 073704
\bibitem{OkuboKawamura-vortex} Okubo T and Kawamura H 2010 J. Phys. Soc. Jpn. {\bf 79} 084706

\bibitem{Olariu} Olariu A, Mendels P, Bert F, Ueland B G, Schiffer P, Berger R F and  Cava R J 2006  Phys. Rev. Lett. {\bf 97} 167203
\bibitem{Hsieh}  Hsieh D, Qian D, Berger R F, Cava R J, Lynn J W, Huang Q, and Hasan M Z 2008 Physica. B {\bf 403} 1341
\bibitem{Hsieh2}  Hsieh D, Qian D, Berger R F, Cava R J, Lynn J W, Huang Q and Hasan M Z 2008 J. Phys. Chem. Solids {\bf 69} 3174
\bibitem{Hemmida} Hemmida M,  Krug von Nidda H A, B\"{u}ttgen N, Loidl A, Alexander L K, Nath R, Mahajan A V, Berger R F, Cava R J, Singh Y, and Johnston D C 2009 Phys. Rev. B {\bf 80} 054406


\bibitem{Nakatsuji} Nakatsuji S, Nambu Y, Tonomura H, Sakai O, Jonas S, Broholm C, Tsunetsugu H, Qiu Y, and Maeno Y 2005 Science {\bf 309} 1697
\bibitem{Nambu} Nambu Y, Nakatsuji S and Maeno Y 2006 J. Phys. Soc. Jpn. {\bf 75} 043711
\bibitem{Takeya} Takeya H, Ishida K, Kitagawa K, Ihara Y, Onuma K, Maeno Y, Nambu Y, Nakatsuji S, MacLaughlin D E, Koda A and Kadono R 2008 Phys. Rev. B {\bf 77} 054429
\bibitem{Yaouanc} Yaouanc A, de R\'{e}otier P D, Chapuis Y, Marin C, Lapertot G, Cervellino A, and Amato A 2008 Phys. Rev. B {\bf 77}  092403
\bibitem{MacLaughlin} MacLaughlin D E, Nambu Y, Nakatsuji S, Heffner R H, Shu L, Bernal O O and Ishida K 2008  Phys. Rev. B {\bf 78} 220403(R)
\bibitem{Yamaguchi} Yamaguchi H, Kimura S, Hagiwara M, Nambu Y, Nakatsuji S, Maeno Y, and Kindo K 2008 Phys. Rev. B {\bf 78} 180404
\bibitem{Yamaguchi2} Yamaguchi H, Kimura S, Hagiwara M, Nambu Y, Nakatsuji S, Maeno Y, Matsuo A and Kindo K 2010 J. Phys. Soc. Jpn. {\bf 79} 054710
\bibitem{Nakatsuji-review} Nakatsuji S, Nambu Y and Onoda S 2010 J. Phys. Soc. Jpn. {\bf 79} 011003
\bibitem{Nambu2} Nambu Y, Nakatsuji S, Maeno Y, Okudzeto E K and Chan J Y 2008 Phys. Rev. Lett. {\bf 101} 207204
\bibitem{Stock} Stock C, Jonas S, Broholm C, Nakatsuji S, Nambu Y, Onuma K, Maeno Y and Chung J H 2010 Phys. Rev. Lett. {\bf 105} 037402
\bibitem{Takubo} Takubo K, Nambu Y, Nakatsuji S, Wakisaka Y, Sudayama T, Fournier D, Levy G, Damascelli A, Arita M, Namatame H, Taniguchi M and Mizokawa T 2010 Phys. Rev. Lett. {\bf 104} 226404

\bibitem{Shimizu} Shimizu Y,  Miyagawa K, Kanoda K,  Maesato M and Saito G 2003 Phys. Rev. Lett. {\bf 91} 107001
\bibitem{Kawamoto} Kawamoto A, Honma Y and Kumagai K 2004 Phys. Rev. B {\bf 70}  R060510
\bibitem{Kurosaki} Kurosaki Y, Shimizu Y, Miyagawa k,  Kanoda K and Saito G 2005 Phys. Rev. Lett. {\bf 95} 177001
\bibitem{Shimizu2} Shimizu Y,  Miyagawa K, Kanoda K,  Maesato M and
      Saito G 2006 Phys. Rev. B {\bf 73} 140407
\bibitem{Ohira} Ohira S, Shimizu Y,  Kanoda K and Saito G 2006 J. Low Temp. Phys. {\bf 142} 153
\bibitem{Yamashita} Yamashita S, Nakazawa Y, Oguni M, Oshima Y, Nojiri H, Shimizu Y, Miyagawa K and Kanoda K 2008 Nature Physics {\bf 4} 459
\bibitem{Yamashita2} Yamashita M, Nakata N, Kasahara Y, Sasaki T, Yoneyama M,  Kobayashi N, Fujimoto S, Shibauchi T and  Matsuda Y 2009 Nature Physics {\bf 5} 44
\bibitem{Manna} Manna R S, de Souza M, Bruhl A, Schlueter J A and Lang M 2010 Phys. Rev. Lett. {\bf 104} 016403
\bibitem{Jawad} Abdel-Jawad M, Terasaki I, Sasaki T, Yoneyama N, Kobayashi N, Uesu Y and Hotta C 2010 Phys. Rev. B {\bf 82} 125119

\bibitem{Tamura} Tamura M,  Nakao A and  Kato R 2006  J. Phys. Soc. Jpn. {\bf 75}  093701
\bibitem{Itou} Itou T, Oyamada A, Maegawa S, Tamura M  and Kato R2008  Phys. Rev. B {\bf 77}  104413
\bibitem{Itou2} Itou T, Oyamada A, Maegawa S and Kato R 2010 Nature Physics {\bf 6} (2010) 673
\bibitem{Yamashita2-dmit} Yamashita M, Nakata N, Senshu Y, Nagata M, Yamamoto H M, Kato R, Shibauchi T and Matsuda Y 2010 Science {\bf 328} 1246

\bibitem{KT1} Kosterliz J M and Thouless D J 1973, J. Phys. C: Solid State Phys. {\bf 6} 1181
\bibitem{KT2} Kosterliz J M and Thouless D J 1974 J. Phys. C: Solid State Phys. {\bf 7} 1046 
\bibitem{APYoung} Southern B W and Young A P 1993 Phys. Rev. B {\bf 48} 13170



\bibitem{Ajiro} Ajiro Y, Kikuchi H, Sugiyama S, Nakashima T, Shamoto S, Nakayama N, Kiyama M, Yamamoto N and Oka Y 1988 J. Phys. Soc. Jpn. {\bf 57} 2268
\bibitem{Soubeyroux} Soubeyroux J L, Fruchart D, Delmas C and Le Flem G 1979  J. Mag. Mag. Mater. {\bf 14} 159

\bibitem{Kadowaki-LiCrO2} Kadowaki H, Takei H and Motoya K 1995 J. Phys. Condens. Matter {\bf 7} 6869
\bibitem{Alexander} Alexander L K, Buttgen N, Nath R, Mahajan A V and Loidl A 2007  Phys. Rev. B {\bf 76} 064429
\bibitem{Tokura} Seki S, Onose Y and Tokura Y 2008 Phys. Rev. Lett. {\bf 101} 067204
\bibitem{Olariu-LiCrO2} Olariu A, Mendels P, Bert F,  Alexander L K, Mahajan A V, Hiller A D and Amato A 2009  Phys. Rev. B {\bf 79} 224401

\bibitem{Kadowaki} Kadowaki H, Kikuchi H and Ajiro Y 1990 J. Phys. Condens. Matter {\bf 2} 4485
\bibitem{Poienar} Poienar M, Damay F,Martin C, Hardy V, Maignan A and Andre G 2009  Phys. Rev. B {\bf 79} 014412
\bibitem{Kimura} Kimura K,  Nakamura H, Kimura S, Hagiwara M and Kimura T 2009 Phys. Rev. Lett. {\bf 103} 107201

\bibitem{Oohara} Oohara Y, Mitsuda S, Yoshizawa H, Yaguchi N, Kuriyama H, Asano T and Mekata M 1994 J. Phys. Soc. Jpn. {\bf 63} 847

\bibitem{Mekata-PdCrO2} Mekata M, Sugino T, Oohara A, Oohara Y and Yoshizawa H 1995 Physica B {\bf 213\& 214} 221
\bibitem{Takatsu} Takatsu H, Yonezawa S, Fujimoto S and Maeno Y 2010 Phys. Rev. Lett. {\bf 105} 137201

\bibitem{Okuda} Okuda T, Kishimoto T, Uto K, Hokazono T, Onose Y, Tokura Y, Kajimoto R and Matsuda M 2009 J. Phys. Soc. Jpn. {\bf 78} 013604
\bibitem{Kajimoto} Kajimoto R, Nakajima K, Ohira-Kawamura S, InamuraY, Kakurai K, Arai M, Hokazono T, Oozono A and Okuda T 2010 J. Phys. Soc. Jpn. {\bf 79} 123705

\bibitem{Fujimoto} Fujimoto S 2006 Phys. Rev. B {\bf 73} 184401
\bibitem{Balents2} Stoudenmire E M, Trebst S and Balents L 2009 Phys. Rev. B {\bf 79} 214436
\bibitem{KawaMiya2} Kawamura H and Miyashita S 1985 J. Phys. Soc. Jpn. {\bf 54} 4530

\bibitem{PLee-spinon1} Lee S S, Lee P A and Senthil T 2007 Phys. Rev. Lett. {\bf 98} 067006
\bibitem{PLee-spinon2} Nave C P and Lee P A 2007 Phys. Rev. B {\bf 76} 235124




\bibitem{H-Yoshida} Yoshida H, Okamoto Y, Tayama T, Sakakibara T, Tokunaga M, Matsuo A, Narumi Y, Kindo K, Yoshida M, Takigawa M and Hiroi Z 2009 J. Phys. Soc. Jpn. {\bf 78} 043704
\bibitem{M-Yoshida} Yoshida M, Takigawa M, Yoshida H, Okamoto Y and Hiroi Z 2009  Phys. Rev. Lett. {\bf 103} 077207
\bibitem{Yamashita-volbo} Yamashita S, Moriura T, Nakazawa Y, Yoshida H, Okamoto Y and Hiroi Z 2010  J. Phys. Soc. Jpn. {\bf 79} 083710
\bibitem{Nilsen} Nilsen G J, Coomer F C, de Vries M A, Stewart J R, Deen P P, Harrison A and Ronnow H M 2010 arXiv:1001.2462
\bibitem{Hiroi} Hiroi Z, Hanawa M, Kobayashi N, Nohara M, Takagi H, Kato Y and Takigawa M 2001  J. Phys. Soc. Jpn. {\bf 70} 3377
\bibitem{Fukaya} Fukaya A, Fukamoto Y, Ito T, Larkin M I, Savici A T, Uemura Y J, Kyriakou P P, Luke G M, Rovers M T, Kojima K M, Keren A, Hanawa M and Hiroi Z 2003  Phys. Rev. Lett. {\bf 91} 207603
\bibitem{BertBono}
Bert F, Bono D, Mendels P, Ladieu F, Duc F, Trombe J-C and Millet P 2005 Phys. Rev. Lett. {\bf 95} 087203
\bibitem{BertBono2}
Bert F, Bono D, Mendels P, Trombe J-C, Millet, Amato A, Baines C and Hiller A 2004 J. Phys. Condens. Matter {\bf 16} S829

\bibitem{Okamoto} Okamoto Y, Yoshida H and Hiroi Z 2009  J. Phys. Soc. Jpn. {\bf 78} 033701
\bibitem{Zhang} Zhang W, Ohta H, Okubo S, Fujisawa M, Sakurai T, Okamoto Y, Yoshida H and Hiroi Z 2010  J. Phys. Soc. Jpn. {\bf 79} 023708
\bibitem{Hiroi-pc} Hiroi Z, private communication

\bibitem{Ramirez}
Ramirez A P, Espinosa G P and Cooper A S 1990 Phys. Rev. Lett. {\bf 64} 2070
; 1992 Phys. Rev. B {\bf 45} 2505
\bibitem{Broholm}
Broholm C, Aeppli G, Espinosa G P and Cooper A S 1990 Phys. Rev. Lett. {\bf 65} 3173
\bibitem{Martinez} Martinez B, 1992 Phys. Rev. B {\bf 46} 10786
\bibitem{Martinez2} Martinez B, Labarta A, Rodriguez-Sol\'{a} R and Obradors X 1994 Phys. Rev. B {\bf 50} 15779
\bibitem{Uemura} Uemura Y J {\it et al.} 1994 Phys. Rev. Lett. {\bf 73} 3306
\bibitem{Lee} Lee S H, Broholm C, Aeppli G, Perring T G, Hessen B and Taylor A 1996 Phys. Rev. Lett. {\bf 76}  4424
\bibitem{Keren} Keren A, Mendels P, Horvatic M, Ferrer F, Uemura Y J, Mekata M and Asano T 1998 Phys. Rev. B{\bf 57} 10745
\bibitem{Ramirez2} Ramirez A P, Hessen B and Winklemann M 2000 Phys. Rev. Lett. {\bf 84} 2957
\bibitem{Mendels} Mendels P, Keren A, Limot L, Mekata M, Collin G and Horvatic M 2000 Phys. Rev. Lett. {\bf 85} 3496
\bibitem{Mekata} Mekata M and Yamada Y 2001 Can J. Phys. {\bf 79} 1421
\bibitem{Limot} Limot L, Mendels P, Collin G, Mondelli C, Ouladdiaf B, Mutka H, Blanchard N and Mekata M 2002 Phys. Rev. B {\bf 65} 144447
\bibitem{KawamuraArimori} Kawamura H and Arimori Y 2002 Phys. Rev. Lett. {\bf 88} 077202
\bibitem{ArimoriKawamura} Arimori Y and Kawamura H 2001 J. Phys. Soc. Jpn. {\bf 70} 3695

\bibitem{Hagemann}
Hagemann I S, Huang Q, Gao X P A, Ramirez A P and Cava R J 2001 Phys. Rev. Lett. {\bf 86} 894
\bibitem{Bono}
Bono D, Mendels P, Collin G and Blanchard N 2004 Phys. Rev. Lett. {\bf 92} 217202
\bibitem{Bono2}
Bono D, Mendels P, Collin G, Blanchard N, Bert F, Amoto A, Bines C and Hiller A D 2004 Phys. Rev. Lett. {\bf 93} 187201
\bibitem{Bonnet}
Bonnet P, Payen C, Mutka H, Danot M, Fabritchnyi P, Stewart J R, Mellergard A and Ritter C 2004 J. Phys. Condens. Matter {\bf 16} S835



























\end{thebibliography}
\end{document}